\documentclass[aps,eqsecnum,preprint,floats,epsf,epsfig,nofootinbib,letter]{revtex4}
\usepackage{epsfig}
\usepackage{multirow}
\usepackage{subfigure}

\begin{document}
\def\be{\begin{eqnarray}}
\def\en{\end{eqnarray}}
\def\non{\nonumber}
\def\la{\langle}
\def\ra{\rangle}
\def\T{{\cal T}}
\def\O{{\cal O}}
\def\B{{\cal B}}
\def\ep{\varepsilon}
\def\hep{\hat{\varepsilon}}
\def\ek{{\vec{\ep}_\perp\cdot\vec{k}_\perp}}
\def\epp{{\vec{\ep}_\perp\cdot\vec{P}_\perp}}
\def\kp{{\vec{k}_\perp\cdot\vec{P}_\perp}}
\def\lsim{ {\ \lower-1.2pt\vbox{\hbox{\rlap{$<$}\lower5pt\vbox{\hbox{$\sim$}
}}}\ } }
\def\gsim{ {\ \lower-1.2pt\vbox{\hbox{\rlap{$>$}\lower5pt\vbox{\hbox{$\sim$}
}}}\ } }
\def\dk{\partial\!\cdot\!K}
\def\pr{{\sl Phys. Rev.}~}
\def\prl{{\sl Phys. Rev. Lett.}~}
\def\pl{{\sl Phys. Lett.}~}
\def\np{{\sl Nucl. Phys.}~}
\def\zp{{\sl Z. Phys.}~}

\font\el=cmbx10 scaled \magstep2{\obeylines\hfill November, 2018}

\vskip 1.5 cm

\centerline{\large\bf Lifetimes of Doubly Charmed Baryons}
\bigskip
\centerline{\bf Hai-Yang Cheng$^a$, Yan-Liang Shi$^b$\footnote{Current address: Cold Spring Harbor Laboratory, 1 Bungtown Road, Cold Spring Harbor, New York 11724, USA}}
\medskip
\centerline{$^a$Institute of Physics, Academia Sinica}
\centerline{Taipei, Taiwan 115, Republic of China}
\medskip
\medskip
\centerline{$^b$C.N. Yang Institute for Theoretical Physics, Stony Brook University} \centerline{Stony Brook, New York 11794, USA}
\bigskip
\bigskip
\bigskip
\bigskip
\bigskip
\centerline{\bf Abstract}
\bigskip
\small
The lifetimes of doubly charmed hadrons are analyzed within the framework of the heavy quark expansion (HQE). Lifetime differences arise from spectator effects such as $W$-exchange and Pauli interference.  The $\Xi_{cc}^{++}$ baryon is longest-lived in the doubly charmed baryon system owing to the destructive Pauli interference absent in the $\Xi_{cc}^+$ and $\Omega_{cc}^+$. In the presence of dimension-7 contributions, its lifetime is reduced from  $\sim5.2\times 10^{-13}s$ to $\sim3.0\times 10^{-13}s$.
The $\Xi_{cc}^{+}$ baryon has the shortest lifetime of order $0.45\times 10^{-13}s$ due to a large contribution from the $W$-exchange box diagram.
It is difficult to make a precise quantitative statement on the lifetime of $\Omega_{cc}^+$. Contrary to $\Xi_{cc}$ baryons, $\tau(\Omega_{cc}^+)$ becomes longer in the presence of dimension-7 effects and the Pauli interference $\Gamma^{\rm int}_+$ even becomes negative. This implies that the subleading corrections are too large to justify the validity of the HQE. Demanding the rate $\Gamma^{\rm int}_+$ to be positive for a sensible HQE, we conjecture that the $\Omega_c^0$ lifetime lies in the range of $(0.75\sim 1.80)\times 10^{-13}s$. The lifetime hierarchy pattern is  $\tau(\Xi_{cc}^{++})>\tau(\Omega_{cc}^+)>\tau(\Xi_{cc}^+)$ and the lifetime ratio $\tau(\Xi_{cc}^{++})/\tau(\Xi_{cc}^+)$ is predicted to be of order 6.7\,.

\pagebreak

\section{Introduction}
Recently, the LHCb collaboration observed a resonance in the $\Lambda_c^+K^-\pi^+\pi^+$ mass spectrum at a mass of $3621.40\pm0.78$ MeV \cite{LHCb:Xiccpp}, which is consistent with expectations for the doubly charmed baryon $\Xi_{cc}^{++}$ baryon.
Subsequently, LHCb presented the first lifetime measurement of this charmed doubly baryon  \cite{LHCb:tauXiccpp}
\be \label{eq:LHCbtauXiccpp}
\tau(\Xi_{cc}^{++})=
(2.56^{+0.24}_{-0.22}\pm0.14)\times 10^{-13}s.
\en
Theoretical predictions available in the literature \cite{Kiselev:1999,Kiselev:2002,Guberina,Chang,Karliner:2014} listed in Table \ref{tab:lifetimes_dc} spread a large range, for example, $\tau(\Xi_{cc}^{++})$ ranges from 0.2 to 1.6 ps. The lifetime hierarchy was predicted to be of the pattern $\tau(\Xi_{cc}^{++})>\tau(\Omega_{cc}^+)>\tau(\Xi_{cc}^+)$ in \cite{Kiselev:2002,Guberina}, but $\tau(\Xi_{cc}^{++})>\tau(\Xi_{cc}^+)>\tau(\Omega_{cc}^+)$ in \cite{Chang}.

In \cite{Cheng:2018} we have shown that the heavy quark expansion (HQE) in $1/m_b$ works well for bottom hadrons. The calculated $B$ meson lifetime ratios $\tau(B^+)/\tau(B^0_d)$ and $\tau(B^0_s)/\tau(B^0_d)$ in HQE are in excellent agreement with experiment, and the computed lifetime ratios
$\tau(\Xi_b^-)/\tau(\Lambda_b^0)$, $\tau(\Xi_b^-)/\tau(\Xi_b^0)$ and $\tau(\Omega_b^-)/\tau(\Xi_b^-)$  also agree well with the data.
 On the contrary, the HQE to $1/m_c^3$ fails to give a satisfactory description of the lifetimes of both charmed mesons and charmed baryons. The HQE to order $1/m_c^3$ implies the lifetime hierarchy $\tau(\Xi_c^+)>\tau(\Lambda_c)>\tau(\Xi_c^0)
>\tau(\Omega_c)$, which seems to be in agreement with the current one from the Particle Data Group (PDG)  \cite{PDG}.
However, the
quantitative estimates of charmed baryon lifetimes and their ratios are
rather poor.  For example, $\tau(\Xi_c^+)/\tau(\Lambda_c^+)$ is calculated to be 1.03 \cite{Cheng:2018}, while experimentally it is measured to be $2.21\pm0.15$ \cite{PDG}. Since the charm quark is not heavy,
it is thus natural to consider the effects stemming from the next-order $1/m_c$ expansion. This calls for the subleading $1/m_Q$ corrections to spectator effects.

It turns out that the relevant dimension-7 spectator effects are in the right direction for explaining the large lifetime ratio of $\tau(\Xi_c^+)/\tau(\Lambda_c^+)$, which is enhanced from 1.05 to 1.88, in better agreement with the experimental value \cite{Cheng:2018}.
However, the destructive $1/m_c$ corrections to $\Gamma(\Omega_c^0)$ are too large to justify the use of the HQE, namely, the predicted Pauli interference and semileptonic rates for the $\Omega_c^0$ become negative, which certainly do not make sense.
Demanding these rates to be positive for a sensible HQE, it has been conjectured in \cite{Cheng:2018} that the $\Omega_c^0$ lifetime lies in the range of $(2.3\sim3.2)\times 10^{-13}s$. This leads to the new lifetime pattern
$\tau(\Xi_c^+)>\tau(\Omega_c^0)>\tau(\Lambda_c^+)>\tau(\Xi_c^0)$, contrary to the current hierarchy $\tau(\Xi_c^+)>\tau(\Lambda_c^+)>\tau(\Xi_c^0)>\tau(\Omega_c^0)$. This new charmed baryon lifetime pattern can be tested by LHCb.

Very recently, LHCb has reported a new measurement of the $\Omega_c^0$ lifetime, $\tau(\Omega_c^0)=(2.68\pm0.24\pm0.10\pm0.02)\times 10^{-13}s$ \cite{LHCb:Omegac}, using the semileptonic decay $\Omega_b^-\to\Omega_c^0\mu^-\bar \nu_\mu X$ with $\Omega_c^0\to pK^- K^-\pi^+$. This value is nearly four times larger than the current world-average value of $\tau(\Omega_c^0)=(0.69\pm0.12)\times 10^{-13}s$ \cite{PDG} from fixed target experiments.
\footnote{
An early conjecture of $\tau(\Omega_c^0)$ of order $2.3\times 10^{-13}s$ first presented  in \cite{Cheng:HIEPA} by one of us is indeed consistent with the LHCb measurement.}
This indicates that
the $\Omega_c^0$, which is naively expected to be shortest-lived in the charmed baryon system owing to the large constructive Pauli interference, could live longer than the $\Lambda_c^+$  due to the suppression from $1/m_c$ corrections arising from dimension-7 four-quark operators.

In this work we shall study the lifetimes of doubly charmed baryons within the framework of the HQE. It is organized as follows. In Sec.~II we give the
general HQE expressions for inclusive nonleptonic and
semileptonic widths. A special attention is paid to the doubly charmed baryon matrix elements of dimension-3 and -5 operators which are somewhat different from the ones of singly charmed baryons. We then proceed to discuss the relevant dimension-6 and -7 four-quark operators. Evaluation of doubly charmed baryon matrix elements and numerical results are presented in Sec. III. Conclusions are given in Sec.~IV.

\begin{table}[t]
\caption{Predicted lifetimes of doubly charmed baryons
in units of $10^{-13}s$. } \label{tab:lifetimes_dc}
\begin{center}
\begin{tabular}{|c|c c c c c |} \hline \hline
 & ~~~Kiselev et al.~~ & ~~~Kiselev et al.~~ & ~~Guberina et al.~~ & ~~Chang et al.~~ & ~~Karliner et al.~~  \\
 &  \cite{Kiselev:1999} &  \cite{Kiselev:2002} & \cite{Guberina} & \cite{Chang} & \cite{Karliner:2014} \\
\hline
 ~~$\Xi_{cc}^{++}$~~~ & $4.3\pm1.1$ & $4.6\pm0.5$ & 15.5 & 6.7 & 1.85 \\
 ~~$\Xi_{cc}^{+}$~~~ & $1.1\pm0.3$ & $1.6\pm0.5$ & 2.2 & 2.5 & 0.53 \\
 ~~$\Omega_{cc}^{+}$~~~ & & $2.7\pm0.6$ & 2.5 & 2.1 &  \\ \hline
 \hline
\end{tabular}
\end{center}
\end{table}

\section{Theoretical framework}
Under the heavy quark expansion,
the inclusive
nonleptonic decay rate of a doubly heavy baryon $\B_{QQ}$ containing two heavy quarks $QQ$
is given by \cite{Bigi92,BS93}
\be
\Gamma(\B_{QQ})={1\over 2m_{\B_{QQ}}}{\rm Im}\,\la \B_{QQ}|T|\B_{QQ}\ra={1\over 2m_{\B_{QQ}}}\la \B_{QQ}|\int d^4x\,T[{\cal L}^\dagger_W(x){\cal L}_W(0)]|\B_{QQ}\ra,
\en
in analog to the case of a singly heavy hadron $H_Q$.
Through the use of the operator product expansion, the transition operator $T$ can be expressed in terms of local quark operators
\be \label{eq:HQE}
{\rm Im}\,T={G_F^2m_Q^5\over 192\pi^3}\,\xi\,\left(c_{3,Q}\bar QQ+{c_{5,Q}\over m_Q^2}\bar Q\sigma\cdot G Q+{c_{6,Q}\over m_Q^3}T_6+ {c_{7,Q}\over m_Q^4}T_7+\cdots\right),
\en
where $\xi$ is the relevant CKM matrix element,
the dimension-6 $T_6$ consists of the four-quark operators $(\bar Q\Gamma q)(\bar q\Gamma Q)$ with $\Gamma$ representing a combination of the Lorentz and color matrices, while a subset of dimension-7 $T_7$ is governed by the four-quark operators containing derivative insertions. Hence,
\be \label{eq:NLrate}
\Gamma(\B_{QQ}) &=& {G_F^2m_Q^5\over 192\pi^3}\,\xi\,{1\over 2m_{\B_{QQ}}}
\Bigg\{ c_{3,Q}\la \B_{QQ}|\bar QQ|\B_{QQ}\ra+ {c_{5,Q}\over m_Q^2} \la\B_{QQ}|\bar Q \sigma\cdot GQ|\B_{QQ}\ra \non \\ &+& {c_{6,Q} \over m_Q^3} \la \B_{QQ}|T_6|\B_{QQ}\ra+{c_{7,Q} \over m_Q^4} \la \B_{QQ}|T_7|\B_{QQ}\ra+\cdots\Bigg\}.
\en

\subsection{Dimension-3 and -5 operators}

In heavy quark effective theory (HQET), the dimension-3 operator $\bar QQ$ in the rest frame has the expression
\be
\bar QQ=\bar Q\gamma_0 Q-{\bar Q(i\vec{D})^2Q\over 2m_Q^2}+{\bar Q\sigma\cdot G Q\over 4m_Q^2}+{\cal O}\left({1\over m_Q^3}\right),
\en
with the normalization
\be
{\la \B_{QQ}|\bar Q\gamma_0 Q|\B_{QQ}\ra \over 2m_{\B_{QQ}}}=1.
\en
Hence,
\be \label{eq:dim3me}
{\la \B_{QQ}|\bar QQ|\B_{QQ}\ra\over 2m_{\B_{QQ}}}=1-{\mu_\pi^2\over 2m_Q^2}+{\mu_G^2\over 2m_Q^2}+{\cal O}\left({1\over m_Q^3}\right),
\en
where
\be \label{eq:lambda12}
&& \mu_\pi^2\equiv {1\over 2m_{\B_{QQ}}}\la \B_{QQ}|\bar Q(i\vec{D})^2Q|\B_{QQ}\ra=-{1\over 2m_{\B_{QQ}}}\la \B_{QQ}|\bar Q(i{D_\perp})^2Q|\B_{QQ}\ra=-\lambda_1,   \non \\
&& \mu_G^2\equiv{1\over 2m_{\B_{QQ}}}\la \B_{QQ}|\bar Q{1\over 2}\sigma\cdot
GQ|\B_{QQ}\ra=d_H\lambda_2.
\en
The non-perturbative parameters $\lambda_1$ and $\lambda_2$ are independent of $m_Q$ and have the same values for all particles in a given spin-flavor multiplet.

We first consider the non-perturbative parameter $\mu_\pi^2$. In general, $\mu_\pi^2=\la p^2\ra=\la m_Q^2v_Q^2\ra$. The average kinetic energy of the diquark $QQ$ and the light quark $q$ is $T={1\over 2}m_dv_d^2+{1\over 2}m_qv_q^2$, where $m_d$ ($m_q$) is the mass of the diquark (light quark). This together with the momentum conservation $m_dv_d=m_qv_q$ leads to
\be
v_d^2={m_q T\over 2m_Q^2+m_Qm_q}.
\en
As shown in \cite{Kiselev:1999}, the average kinetic energy $T'$ of heavy quarks inside the diquark given by ${1\over 2}m_Q(v_{Q1}^2+v_{Q2}^2)$ is equal to $T/2$ due to the color wave function of the diquark. Hence, the average velocity  $\tilde v$ of the heavy quark inside the diquark is $\tilde v^2=T/(2m_Q)$. The average velocity $v_Q$ of the heavy quark inside the baryon $\B_{QQ}$ is \cite{Kiselev:1999}
\be
v_Q^2\approx \tilde v^2+v_d^2={T\over 2m_Q}+{m_q T\over 2m_Q^2+m_Qm_q}.
\en
Hence,
\be
\mu_\pi^2(\B_{QQ})\simeq m_Q\left( {T\over 2}+{m_q T\over 2m_Q+m_q}\right).
\en

We next turn to the parameter $\mu_G^2$. In HQET,  the mass of the singly heavy baryon  $\B_{Q}$ has the expression
\be \label{eq:mHQ}
m_{\B_{Q}}=\,m_Q+\bar\Lambda_{\B_{Q}}+{\mu_\pi^2\over 2m_Q}-{\mu_G^2\over 2m_Q}+{\cal O}\left({1\over m_Q^2}\right),
\en
where $\bar\Lambda_{\B_{Q}}$ is a parameter of HQET and it can be regarded as the binding energy of the heavy hadron in the infinite mass limit. For the doubly heavy baryon $\B_{QQ}$, if the heavy diquark acts as a point-like constitute, its mass is  of the form
\be \label{eq:mHQQ}
m_{\B_{QQ}}=\,2m_Q+\bar\Lambda_{\B_{QQ}}+{\mu_\pi^2\over m_Q}-{\mu_G^2\over m_Q}+{\cal O}\left({1\over m_Q^2}\right).
\en
There are two distinct chromomagnetic fields inside the $\B_{QQ}$:
one is  the chromomagnetic field produced by the light quark and the other
by the heavy quark. For the former (latter), the operator $\sigma\cdot G$ is proportional to $\vec{S}_d\cdot\vec{S}_q$ ($\vec{S}_1\cdot\vec{S}_2$),  where $\vec{S}_d=\vec{S}_1+\vec{S}_2$ ($\vec{S}_q$) is the spin
operator of the diquark (light quark), and $\vec{S}_i$ ($i=1,2$) is the spin of the constituent quark inside the diquark.
The parameter $d_H$ is given by
\footnote{The coefficients of $\vec{S}_d\cdot\vec{S}_q$ and $\vec{S}_1\cdot\vec{S}_2$ can be arbitrarily chosen. The $\mu_G^2$ term is independent of the choice of $d_H$.}
\be
d_H^{dq} &=& -4\la \B_{QQ}|\vec{S}_d\cdot\vec{S}_q|\B_{QQ}\ra=
 -2[S_{\rm tot}(S_{\rm tot}+1)-S_d(S_d+1)-S_q(S_q+1)],  \non \\
d_H^{QQ} &=& -4\la \B_{QQ}|\vec{S}_1\cdot\vec{S}_2|\B_{QQ}\ra
 =-2[S_d(S_d+1)-S_1(S_1+1)-S_2(S_2+1)].
\en
Therefore,  $d_H^{dq}=4$,  $d_H^{QQ}=-1$ for the spin-${1\over 2}$
doubly heavy baryon $\B_{QQ}$ and $d_H^{dq}=-2$,  $d_H^{QQ}=-1$ for the spin-${3\over 2}$ doubly heavy baryon
$\B^*_{QQ}$. It follows from Eq. (\ref{eq:mHQQ}) that $\lambda_2^{dq}$ can be expressed in terms of the hyperfine mass splitting
\be
\lambda_2^{dq}(\B_{QQ}) = {1\over 6}(m_{\B^*_{QQ}}-m_{\B_{QQ}})m_Q,
\en
and hence,
\be
\mu_G^2(\B_{QQ})={2\over 3}(m_{\B^*_{QQ}}-m_{\B_{QQ}})m_Q-\lambda_2^{QQ}.
\en

To evaluate the parameter $\lambda_2^{QQ}$, let us consider a simple quark model of De R\'ujula {\it et al.} \cite{DeRujula}
\be
M_{\rm baryon} &=& M_0+\cdots+{16\over 9}\pi\alpha_s\sum_{i>j}{\vec{S}_i\cdot\vec{S}_j\over m_im_j}|\psi(0)|^2, \non \\
M_{\rm meson} &=& M_0+\cdots+{32\over 9}\pi\alpha_s {\vec{S}_1\cdot\vec{S}_2\over m_1m_2}|\psi(0)|^2.
\en
It is well known that the fine structure constant is $-{4\over 3}\alpha_s$ for $\bar qq$ pairs in a meson and $-{2\over 3}\alpha_s$ for $qq$ pairs in a baryon \cite{DeRujula}. This is because the $\bar qq$ pair in a meson must be a color-singlet, while the $qq$ pair in a baryon is in color antitriplet state.
The mass of the doubly heavy baryon $\B_{QQ}$ is given by
\be
m_{\B_{QQ}}=2m_Q+\cdots+{16\over 9}\pi\alpha_s\left( {\vec{S}_{d}\cdot\vec{S}_q \over m_Qm_q }|\psi^{dq}(0)|^2+{\vec{S}_1\cdot\vec{S}_2\over m_Q^2}|\psi^{QQ}(0)|^2\right)+{\cal O}\left({1\over m_Q^2}\right),
\en
where  $\psi^{dq}(0)$ is the light quark wave function at the origin of the $QQ$ diquark and $\psi^{QQ}(0)$ is the diquark wave function at the origin.
For the doubly charmed baryons we have
\be \label{eq:massXicc}
m_{\Xi_{cc}} &=& 2m_c+\cdots+{16\over 9}\pi\alpha_s\left( -{1\over m_cm_q} |\psi^{dq}(0)|^2+{1\over 4m_c^2}|\psi^{cc}(0)|^2\right)+{\cal O}\left({1\over m_c^2}\right), \non \\
m_{\Xi_{cc}^*} &=& 2m_c+\cdots+{16\over 9}\pi\alpha_s\left( {1\over 2m_cm_q} |\psi^{dq}(0)|^2+{1\over 4m_c^2}|\psi^{cc}(0)|^2\right)+{\cal O}\left({1\over m_c^2}\right).
\en
The term proportional to $|\psi^{dq}(0)|^2$ can be expressed in terms of the hyperfine mass splitting of $\Xi_{cc}$. Hence, we obtain
\be
\mu_G^2(\Xi_{cc})={2\over 3}(m_{\Xi_{cc}^*}-m_{\Xi_{cc}})m_c-{4\over 9}\pi\alpha_s{|\psi^{cc}(0)|^2\over m_c}+{\cal O}\left({1\over m_c}\right).
\en
Hence, $\lambda_2^{cc}(\Xi_{cc})=(1/9)g_s^2{|\psi^{cc}(0)|^2/ m_c}$.

However, the above expression of $\mu_G^2$ is not the end of story.
It has been known that HQET is not the appropriate effective field theory for hadrons with more than one heavy quark. HQET is formulated as an expansion in $\Lambda_{\rm QCD}/m_Q$. For a singly heavy hadron, the heavy quark kinetic energy is neglected as it occurs as a small $1/m_Q$ correction. For a bound state containing two or more heavy quarks, the heavy quark kinetic energy is very important and cannot be treated as a perturbation. The appropriate theory for dealing such a system is non-relativistic QCD (NRQCD),
\footnote{However, it was pointed out very recently in \cite{An} that in the limit $m_Q>m_Qv_Q>m_Qv_Q^2\gg\Lambda_{\rm QCD}$, such a system can be described by a version of HQET with a diquark degree of freedom.}
in which one has
\be
\bar Qg_s\sigma\cdot GQ=-2\psi_Q^\dagger g_s\vec{\sigma}\cdot\vec{B}\psi_Q-{1\over m_Q}\psi^\dagger_Q g_s\vec{D}\cdot \vec{E}\psi_Q+\cdots
\en
in terms of the two-spinor $\psi_Q$. According to the counting rule, the Darwin term for the interaction with the chromoelectric field is of the same order of magnitude as the chromomagnetic term \cite{Beneke:Bc}.  Hence, we get an additional contribution to $\mu_G^2$
\be \label{eq:muG}
\mu_G^2={2\over 3}(m_{\Xi_{cc}^*}-m_{\Xi_{cc}})m_c-{1\over 9}g_s^2{|\psi^{cc}(0)|^2\over m_c}-{1\over 6}g^2_s{|\psi^{cc}(0)|^2\over m_c}.
\en
The last term can be obtained by using the equation of motion for the chromoelectric field.
Note that our result is different from the original expression
\footnote{Guberina {\it et al.} \cite{Guberina} obtained a similar  expression except for the magnitude of $|\psi^{cc}(0)|^2$ terms
\be
\mu_G^2={2\over 3}(m_{\Xi_{cc}^*}-m_{\Xi_{cc}})m_c-{2\over 9}g_s^2{|\psi^{cc}(0)|^2\over m^*_c}-{1\over 3}g_s^2{|\psi^{cc}(0)|^2\over m_c}.\non
\en
}
\be
\mu_G^2=-{2\over 3}(m_{\Xi_{cc}^*}-m_{\Xi_{cc}})m_c-{2\over 9}g_s^2{|\psi^{cc}(0)|^2\over m_c}-{1\over 3}g_s^2{|\psi^{cc}(0)|^2\over m_c}
\en
obtained in \cite{Kiselev:1999} in the sign of the first term and in the magnitude of $|\psi^{cc}(0)|^2$ terms.
Therefore,
\be
{\la \Xi_{cc}|\bar cc|\Xi_{cc}\ra\over 2m_{\Xi_{cc}}}=1-{1\over 2}v_c^2+{1\over 3}{ m_{\Xi_{cc}^*}-m_{\Xi_{cc}}\over m_c} -{2\over 9}\pi\alpha_s{|\psi^{cc}(0)|^2\over m_c^3}-{1\over 3}\pi\alpha_s{|\psi^{cc}(0)|^2\over m_c^3}
\en

Since the hyperfine mass splitting of $D$ mesons is given by
\be
m_{D^*}-m_D={32\over 9}\alpha_s\pi {|\psi^{c\bar q}(0)|^2\over m_cm_q},
\en
we are led to the relation
\be
m_{\Xi^*_{cc}}-m_{\Xi_{cc}}={3\over 4}(m_{D^*}-m_D)\,{|\psi^{dq}_{\Xi_{cc}}(0)|^2\over |\psi^{c\bar q}_{D_q}(0)|^2}.
\en
In the heavy quark limit, the doubly charmed baryon wave function $\psi^{dq}_{\Xi_{cc}}(0)$ is expected to be the same as the meson wave function $\psi^{c\bar q}_{D_q}(0)$ if the diquark behaves as a point-like particle,
\footnote{In \cite{Guberina} and in \cite{Onishchenko}, the authors argued that $|\psi^{dq}(0)|^2={2\over 3}|\psi^{c\bar q}(0)|^2$ due to different spin content of doubly charmed baryons. However, this will not lead to the approximate mass relation given by Eq. (\ref{eq:massrel}).}
\be \label{eq:samew.f.}
 \psi^{dq}_{\Xi_{cc}}(0)\approx \psi^{c\bar q}_{D_q}(0).
 \en
It follows the well-known mass relation
\be \label{eq:massrel}
m_{\Xi^*_{cc}}-m_{\Xi_{cc}}={3\over 4}(m_{D^*}-m_D),
\en
which has been derived in various contents, such as HQET \cite{Savage},
\footnote{A factor of 2 was missed in the original mass relation derived in
\cite{Savage}.}
pNRQCD (potential NRQCD) \cite{Brambilla,Fleming} and the quark model \cite{Lewis,Ebert:2005}.

The nonleptonic and semiletponic decay rates of the heavy quark $c$ of the $\B_{cc}$  are given by
\be \label{eq:dec}
\Gamma^{\rm dec}(\B_{cc}) &=& 2{G_F^2m_c^5\over 192\pi^3}\,\xi\,
\Bigg\{ c_{3,c}^{\rm NL} \Big[1-{\mu_\pi^2\over 2m_c^2}+{\mu_G^2\over 2m_c^2}\Big]+ 2c_{5,c}^{\rm NL}{\mu_G^2\over m_c^2}\Bigg\}
\en
and
\be \label{eq:SLrate}
\Gamma^{\rm SL}(\B_{cc}) &=& 2{G_F^2m_c^5\over 192\pi^3}\,\xi\,
\Bigg\{ c_{3,c}^{\rm SL} \Big[1-{\mu_\pi^2\over 2m_c^2}+{\mu_G^2\over 2m_c^2}\Big]+ 2c_{5,c}^{\rm SL}{\mu^2_G\over m_c^2} \Bigg\},
\en
where the expressions of the coefficients $c_{3,c}$ and $c_{5,c}$ can be found, for example, in \cite{Cheng:2018}.

\begin{figure}[t]
\begin{center}
\subfigure[]{
\includegraphics[height=30mm]{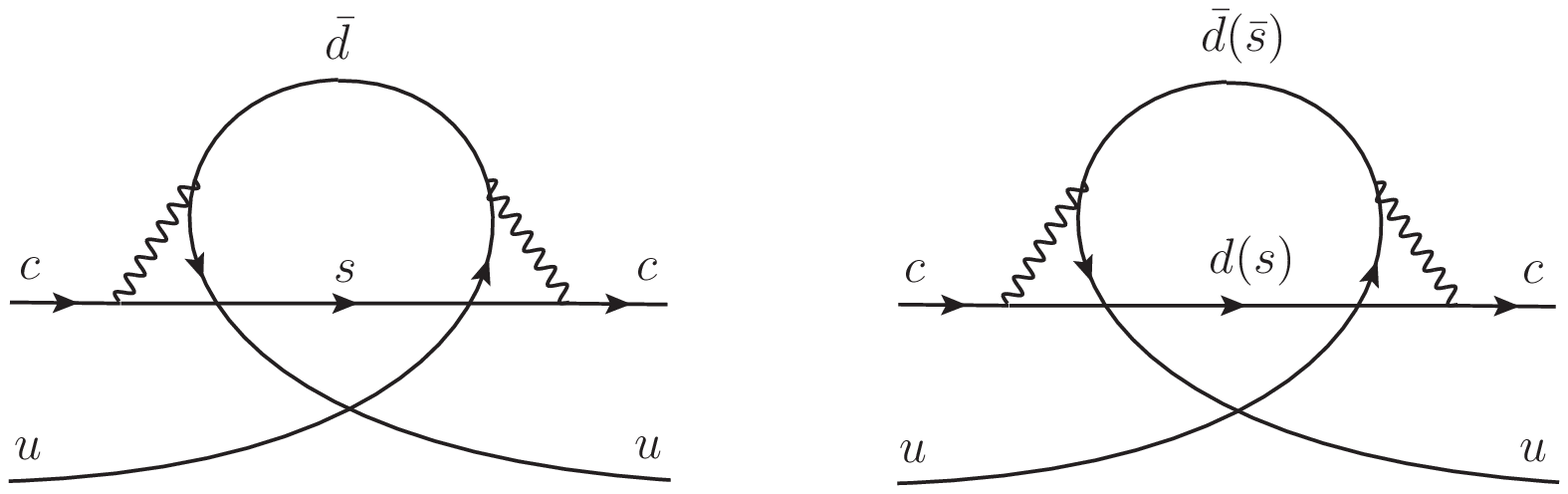}}
\subfigure[]{
\vspace{0.0cm}
\includegraphics[height=20mm]{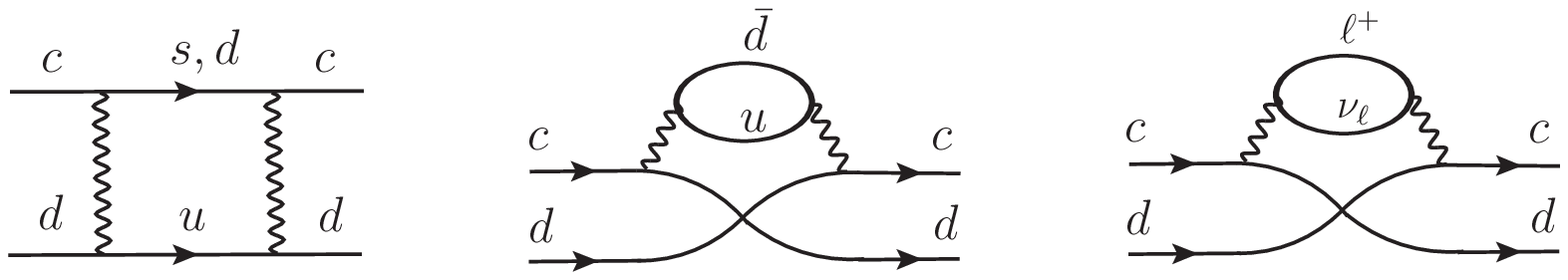}}
\subfigure[]{
\vspace{0.0cm}
\includegraphics[height=20mm]{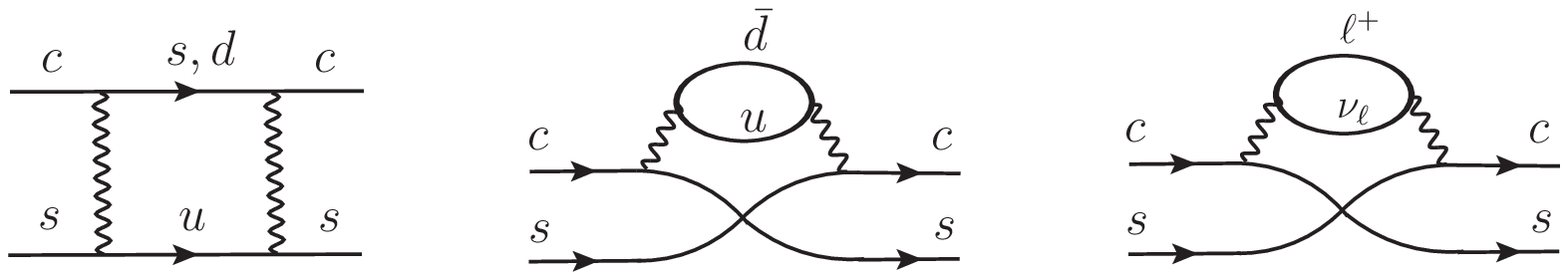}
}
\caption{Spectator effects in doubly charmed baryon decays: (a) destructive Pauli interference in $\Xi_{cc}^{++}$ decay, (b) $W$-exchange and constructive Pauli interference in $\Xi_{cc}^+$ decay, and (c) $W$-exchange and constructive Pauli interference in $\Omega_{cc}^+$ decay.}
\label{fig:spectator}
\end{center}
\end{figure}

\subsection{Dimension-6 operators}

 Defining
\be
\T_6={G_F^2m_Q^2\over 192\pi^3}\xi\,c_{6,Q}^{\rm NL}\,T_6,
\en
the dimension-6 four-quark operators in Eq. (\ref{eq:NLrate}) for spectator effects in inclusive decays of doubly charmed baryons denoted by $\B_{cc}$ are given by (only Cabibbo-allowed decays with $\xi=|V_{cs}V_{ud}|^2$ being listed here)
\cite{Bilic,Guberina:1986,SV}
\be  \label{eq:T6baryon}
\T_{6,ann}^{\B_{cc},d} &=& {G^2_Fm_c^2\over 2\pi}\,\xi\,(1-x)^2\Big\{
(c_1^2+c_2^2)(\bar cc)(\bar dd)+2c_1c_2(\bar cd)(\bar d c)\Big\},
 \non \\
\T_{6,int-}^{\B_{cc},u} &=& -{G_F^2m_c^2\over 6\pi}\,\xi (1-x)^2\Bigg\{ c_1^2\left[ (1+
{x\over 2})(\bar cc)(\bar uu)-(1+2x)\bar c^\alpha(1-\gamma_5)u^\beta
\bar u^\beta(1+\gamma_5)c^\alpha\right]   \non \\
&+& (2c_1c_2+N_cc_2^2)\left[ (1+{x\over 2})(\bar cu)(\bar uc)-
(1+2x)\bar c(1-\gamma_5)u\bar u(1+\gamma_5)c\right]\Bigg\},   \\
\T_{6,int+}^{\B_{cc},s} &=& -{G_F^2m_c^2\over 6\pi}\,\xi\Bigg\{
c_2^2\left[(\bar cc)(\bar ss)-\bar c^\alpha(1-\gamma_5)s^\beta
\bar s^\beta(1+\gamma_5)c^\alpha\right]   \non \\
&+& (2c_1c_2+N_cc_1^2)\Big[ (\bar cs)(\bar sc)-
\bar c(1-\gamma_5)s\bar s(1+\gamma_5)c\Big] \Bigg\}, \non
\en
where $(\bar q_1q_2)\equiv \bar q_1\gamma_\mu(1-\gamma_5)q_2$, and $\alpha,~\beta$
are color indices and $x=m_s^2/m_c^2$.

Spectator effects in the weak decays of the doubly charmed baryons $\Xi_{cc}^{++}$, $\Xi_{cc}^{+}$ and $\Omega_{cc}^+$ are depicted in Fig. \ref{fig:spectator}. The first term $\T_{6,ann}^{\B_{cc},d}$ in (\ref{eq:T6baryon}) corresponds to a $W$-exchange contribution which appears in $\Xi_{cc}^+$ decays (Cabibbo-suppressed $\T_{6,ann}^{\B_{cc},s}$ term appearing in $\Omega_{cc}^+$ decays). The second term $\T_{6,int-}^{\B_{cc},u}$ arises from the destructive Pauli interference of the $u$ quark produced in the $c$ quark decay with the $u$ quark in the wave function of the doubly charm baryon $\B_{cc}$, namely $\Xi_{cc}^{++}$ (Fig. \ref{fig:spectator}(a)). The last term $\T_{6,int+}^{\B_{cc},s}$ is due to the constructive interference of the $s$ quark and hence it occurs only
in charmed baryon decays (Fig. \ref{fig:spectator}(c)).

   For inclusive semileptonic decays, apart from the heavy quark decay
contribution there is an additional spectator effect in charmed-baryon
semileptonic decay originating from the Pauli interference of the $s$ or $d$
quark \cite{Voloshin}; that is, the $s$ ($d$) quark produced in $c\to s\ell^+\nu_\ell$ ($c\to d\ell^+\nu_\ell$) has an interference with the $s$ ($d$) quark in the wave function of the charmed baryon (see Fig. \ref{fig:spectator}).
It is now ready to deduce this term from $\T_{6,int+}^{q_3}$ in Eq. (\ref{eq:T6baryon})  by putting $c_1=1$, $c_2=0$, $N_c=1$:
\be \label{eq:SL6}
\T_{6,int}^{\rm SL} &=&
 -{G_F^2m_c^2\over 6\pi}\Big(|V_{cs}V_{ud}|^2\,
[(\bar c s)(\bar s c)-\bar c(1-\gamma_5)s\bar s(1+\gamma_5)c] \non \\
 && + |V_{cd}V_{ud}|^2\,
[(\bar c d)(\bar d c)-\bar c(1-\gamma_5)d\bar d(1+\gamma_5)c] \Big)
\en

Before proceeding, we would like to clarify how the heavy quark expansion and approximation are consistent with the claimed accuracy. For example, the hadronic matrix element of the dimension-3 operator $\bar QQ$,  Eq. (\ref{eq:dim3me}), is in itself an approximation valid up to
corrections of order $1/m^3_Q$. This is because the chromomagnetic operator $\mu_G^2$ given in Eq. (\ref{eq:muG}), for instance, is valid up to $1/m_Q$ corrections  stemming from  the expansion of Eq. (\ref{eq:mHQQ}) truncated at order $1/m_Q$. Hence,  to the order of $1/m_Q^3$ expansion in Eq. (\ref{eq:NLrate}), one may wonder if it is necessary to take into account the higher order corrections such as $c_{3,Q}\O(1/m_Q^3)+c_{5,Q}\O(1/m_Q^3)$ besides the dimension-6 operator  $c_{6,Q}T_6/m_Q^3$. It turns out that higher order corrections can be neglected as there is a two-body phase-space enhancement factor of $16\pi^2$ for spectator effects induced by dimension-6 four-quark operators $T_6$ relative to the three-body phase space for heavy quark decay. Indeed, the phase-space enhancement for spectator effects is already taken into account in Eq. (\ref{eq:T6baryon}).
Likewise, higher order corrections $c_{3,Q}\O(1/m_Q^4)+c_{5,Q}\O(1/m_Q^4)$ should be less important than the dimension-7 operators $c_{7,Q}T_6/m_Q^4$.

\subsection{Dimension-7 operators}
To the order of $1/m_Q^4$ in the heavy quark expansion in Eq. (\ref{eq:NLrate}), we need to consider dimension-7 operators. For our purposes, we shall focus on the $1/m_Q$
corrections to the spectator effects discussed in the last subsection and neglect the operators with gluon fields.
Dimension-7 terms are either the four-quark operators times the spectator quark mass or the four-quark operators with one or two additional derivatives \cite{Gabbiani:2003pq,Gabbiani:2004tp}.
We shall follow \cite{Lenz:D} to define the following dimension-7 four-quark operators:
\be
&& P_1^q={m_q\over m_Q}\bar Q(1-\gamma_5)q\bar q(1-\gamma_5)Q, \qquad\qquad\qquad
~~P_2^q={m_q\over m_Q}\bar Q(1+\gamma_5)q\bar q(1+\gamma_5)Q, \non \\
&& P_3^q={1\over m_Q^2}\bar Q \stackrel{\leftarrow}{D}_\rho\gamma_\mu(1-\gamma_5)D^\rho q\bar q\gamma^\mu(1-\gamma_5)Q, \quad
~P_4^q={1\over m_Q^2}\bar Q \stackrel{\leftarrow}{D}_\rho(1-\gamma_5)D^\rho q\bar q(1+\gamma_5)Q, \non \\
&& P_5^q={1\over m_Q}\bar Q \gamma_\mu(1-\gamma_5)q\bar q\gamma^\mu(1-\gamma_5)(iD\!\!\!\!/)Q, \quad\quad
~~P_6^q={1\over m_Q}\bar Q (1-\gamma_5)q\bar q(1+\gamma_5)(iD\!\!\!\!/)Q, \non
\en
and the color-octet operators $S_i^q$ ($i=1,...,6$) obtained from $P_i^q$ by inserting $t^a$ in the two currents of the respective color singlet operators.
In order to evaluate the baryon matrix elements, it is more convenient to express dimension-7 operators in terms of $P_i^q$ and $\tilde P_i^q$ operators, where $\tilde P_i$ denotes the color-rearranged operator that follows from the expression of $P_i$ by interchanging the color indices of the $q_i$ and $\bar q_j$ Dirac spinors.  We shall see below that the hadronic matrix elements of dimension-7 operators are suppressed relative to that of dimension-6 ones by order $m_q/m_c$.

Using the relation
\be
S_i=-{1\over 2N_c}P_i+{1\over 2}\tilde P_i,
\en
we obtain \cite{Cheng:2018}
\be \label{eq:T7baryon}
\T_{7,ann}^{\B_{cc},d} &=& {G^2_Fm_c^2\over 2\pi}\,\xi\,(1-x)\Bigg\{ 2c_1c_2\Big[2(1+x)P_3^{d}+(1-x)P^{d}_5\Big] \non \\
&+& (c_1^2+c_2^2)\left[2(1+x)\tilde P_3^{d}+(1-x)\tilde P^{d}_5\right]\Bigg\},
\non\\
\T_{7,int}^{\B_{cc},u} &=& {G_F^2m_c^2\over 6\pi}\,\xi(1-x)\Bigg\{ \Big(2c_1c_2+N_cc_2^2\Big)\Big[-(1-x)(1+2x)(P_1^{u}+P_2^{u}) \non \\
&+& 2(1+x+x^2)P_3^{u}
-12x^2P_4^{u}-(1-x)(1+{x\over 2})P_5^{u}+(1-x)(1+2x)P_6^{u}\Big] \non \\
 &+& c_1^2\Big[ -(1-x)(1+2x)(\tilde P_1^{u}+\tilde P_2^{u})
+ 2(1+x+x^2)\tilde P_3^{u}-12x^2\tilde P_4^{u} \non \\
&-& (1-x)(1+{x\over 2})\tilde P_5^{u}+(1-x)(1+2x)\tilde P_6^{u}\Big]
\Bigg\},   \\
\T_{7,int}^{\B_{cc},s} &=& {G_F^2m_c^2\over 6\pi}\,\xi\Bigg\{ \left(2c_1c_2+N_cc_1^2\right)\Big[-P_1^s-P_2^s
+ 2P_3^s -P_5^s+P_6^s\Big] \non \\
 &+& c_2^2\Big[ -\tilde P_1^s-\tilde P_2^s
+ 2\tilde P_3^s - \tilde P_5^s+\tilde P_6^s\Big]
\Bigg\}. \non
\en
As for the dimension-7 four-operator for semileptonic decays, it can be obtained  from $\T_{7,int}^{\B_{cc},s}$ by setting $c_1=1$, $c_2=0$ and $N_c=1$. Taking into account the lepton mass corrections, it reads \cite{Cheng:2018}
\be  \label{eq:SL7}
\T_{7,int}^{\rm SL} &=& {G_F^2m_c^2\over 6\pi}\,\xi\Big[-(1-z)^2(1+2z)(P_1^s+P_2^s)
+ 2(1-z)(1+z+z^2)P_3^s  \non \\
&& -12z^2(1-z)P_4^s\Big],
\en
where $z=(m_\ell/m_c)^2$.

\section{Lifetimes of doubly charmed baryons}

The inclusive
nonleptonic rates of doubly charmed baryons in the valence quark
approximation and in the limit $m_s/m_c=0$ can be expressed approximately as
 \begin{eqnarray} \label{eq:lifetimes_dc}
 \Gamma_{\rm NL}(\Xi_{cc}^{++}) &=& \Gamma^{\rm
 dec}+\Gamma^{\rm
 int}_-,  \nonumber \\
 \Gamma_{\rm NL}(\Xi_{cc}^{+}) &=& \Gamma^{\rm
 dec}+\cos\theta_C^2\Gamma^{\rm ann}+\sin\theta_C^2\Gamma^{\rm int}_+,  \nonumber \\
 \Gamma_{\rm NL}(\Omega_{cc}^+) &=& \Gamma^{\rm
 dec}+\sin\theta_C^2\Gamma^{\rm ann}+\cos\theta_C^2\Gamma^{\rm int}_+.
 \end{eqnarray}
Because $\Gamma^{\rm int}_+$ is positive and $\Gamma^{\rm int}_-$ is
negative,  it is obvious that $\Xi_{cc}^{++}$ is longest-lived, whereas $\Xi_{cc}^+$ ($\Omega_{cc}^+$) is the shortest-lived if $\Gamma^{\rm int}_+>\Gamma^{\rm ann}$ ($\Gamma^{\rm int}_+<\Gamma^{\rm ann}$). In this section, we shall begin with the evaluation of the doubly charmed baryon matrix elements of dimension-6 and -7 operators and then proceed to compute the spectator effects to see the relative weight between $\Gamma^{\rm int}_+$ and $\Gamma^{\rm ann}$.

\subsection{Baryon matrix elements}

The spectator effects in inclusive heavy bottom baryon decays arising from dimension-6 and dimension-7 operators are given by Eqs. (\ref{eq:T6baryon}), (\ref{eq:SL6}), (\ref{eq:T7baryon}) and (\ref{eq:SL7}), respectively.
We shall rely on the quark model to evaluate the baryon matrix elements of four-quark operators. The $cc$ diquark of the doubly charmed baryon is of the axial-vector type with spin 1. Hence, the $\B_{cc}$ matrix elements of dimension-6 four-quark operators are similar to that of the sextet singly charmed baryon $\Omega_c^0$.
Following \cite{Cheng:2018}, we write down the relevant baryon matrix elements of dimension-6 operators
\be \label{eq:m.e.dim6}
&& \la \B_{cc}|(\bar cq)(\bar qc)|\B_{cc}\ra = -f_{{D_q}}^2m_{_{D_q}}m_{_{\B_{cc}}} r_{_{\B_{cc}}},  \non \\
&& \la \B_{cc}|(\bar cc)(\bar qq)|\B_{cc}\ra = f_{{D_q}}^2m_{_{D_q}}m_{_{\B_{cc}}} r_{_{\B_{cc}}}\tilde B, \non \\
&& \la \B_{cc}|\bar c(1-\gamma_5)q\bar q(1+\gamma_5)c|\B_{cc}\ra = -{1\over 6}f_{{D_q}}^2m_{_{D_q}}m_{_{\B_{cc}}}r_{_{\B_{cc}}}, \\
&& \la \B_{cc}|\bar c^\alpha(1-\gamma_5)q^\beta\bar q^\beta(1+\gamma_5)c^\alpha|
\B_{cc}\ra = {1\over 6}f_{{D_q}}^2m_{_{D_q}}m_{_{\B_{cc}}}r_{_{\B_{cc}}}\tilde B, \non
\en
where $f_{D_q}$ and $m_{_{D_q}}$ are the decay constant and the mass of the heavy meson $D_q$, respectively, and the wave function ratio $r_{\B_{cc}}$ is defined by
\be \label{eq:rb}
r_{\Xi_{cc}}\equiv \left|{\psi_{\Xi_{cc}}^{dq}(0)\over \psi_D^{c\bar q}(0)}\right|^2={4\over 3}{m_{\Xi_{cc}^*}-m_{\Xi_{cc}}\over m_{D^*}-m_D}, \qquad
r_{\Omega_{cc}}\equiv \left|{\psi_{\Omega_{cc}}^{ds}(0)\over \psi_{D_s}^{c\bar s}(0)}\right|^2={4\over 3}{m_{\Omega_{cc}^*}-m_{\Omega_{cc}}\over m_{D_s^*}-m_{D_s}}.
\en
According to Eq. (\ref{eq:samew.f.}), we should have $r_{\Xi_{cc}}=1$.
\footnote{Using the masses of doubly charmed baryons calculated in the relativistic quark model \cite{Ebert:2005}, we find numerically $r_{\Xi_{cc}}=0.99$ and
$r_{\Omega_{cc}}=0.87$.}
The parameter $\tilde B$ is defined by
\be \label{eq:tildeB}
\la \B_{cc}|(\bar cc)(\bar qq)|\B_{cc}\ra=-\tilde B \la \B_{cc}|(\bar cq)(\bar qc)|\B_{cc}\ra.
\en
Since the color wavefunction for a baryon is totally antisymmetric,
the matrix element of $(\bar cc)(\bar qq)$ is the same as that of $(\bar cq)
(\bar qc)$ except for a sign difference. That is, $\tilde B=1$ under the valence-quark approximation.

Likewise,  the $\B_{cc}$ matrix elements of dimension-7 operators are similar to that of the sextet singly charmed baryon $\Omega_c^0$ (see Eq. (4.14) of \cite{Cheng:2018})
\be \label{eq:m.e.dim7}
&& \la \B_{cc}|P_1^q|\B_{cc}\ra=\la \B_{cc}|P_2^q|\B_{cc}\ra={1\over 8} f_{D_q}^2 m_{_{D_q}}m_{_{\B_{cc}}}r_{_{\B_{cc}}}\left( {m^2_{\B_{cc}}-m_{\{cc\}}^2\over m_c^2}\right)\eta^q_{1,2},  \non \\
&& \la \B_{cc}|P_3^q|\B_{cc}\ra= 6\la \B_{cc}|P_4^q|\B_{cc}\ra=-{1\over 4} f_{D_q}^2 m_{_{D_q}}m_{_{\B_{cc}}}r_{_{\B_{cc}}}\left( {m^2_{\B_{cc}}-m^2_{\{cc\}}\over m_c^2}\right)\eta^q_{3,4},
\en
where the parameters $\eta_i^q$ are expected to be of order unity, and $m_{\{cc\}}$  is the mass of the $cc$ diquark.  We take $m_{\{cc\}}$ to be 3226 MeV obtained from the relativistic quark model \cite{Ebert:2005}. Note that the term
\be
{m_{\B_{cc}}^2-m_{\{cc\}}^2\over m_c^2}\approx 4{p_c\cdot p_q\over m_c^2}
\en
is of order $m_q/m_c$. Therefore, the matrix elements of dimension-7 operators are suppressed by a factor of $m_q/m_c$ relative to that of dimension-6 ones.
For the matrix elements of the operators $\tilde P_i^q$,  we introduce a parameter $\tilde\beta_i^q$ in analog to Eq. (\ref{eq:tildeB})
\be \label{eq:beta}
\la \B_{cc}|\tilde P_i^q|\B_{cc}\ra=-\tilde \beta_i^q \la \B_{cc}|P_i^q|\B_{cc}\ra,
\en
so that $\tilde\beta_i^q=1$ under the valence quark approximation.

For the spectator effects in doubly charmed baryon decays,
\be
\Gamma^{\rm spec}(\B_{cc})={\la \B_{cc}|\T_6+\T_7|\B_{cc}\ra\over 2m_{\B_{cc}}},
\en
we apply Eqs. (\ref{eq:m.e.dim6}) and (\ref{eq:m.e.dim7}) to evaluate the matrix elements of the dimension-6 and -7 operators. The results are
\be \label{eq:Spectorcharmbary}
\Gamma^{\rm ann}(\Omega_{cc}^+) &=& 3{G_F^2m_c^2\over \pi}\,|V_{cs}V_{us}|^2\,
r_{{\Omega_{cc}}}\left|\psi^{D_s}_{c\bar s}(0)\right|^2 \left((1-x)^2+\left|{V_{cd}\over V_{cs}}\right|^2\right)\Bigg\{\Big(\tilde B(c_1^2+c_2^2)-2c_1c_2\Big) \non \\
&+& {1\over 2}\Big(\tilde \beta(c_1^2+c_2^2)-2c_1c_2\Big)\eta\left( {m^2_{\Omega_{cc}}-m^2_{\{cc\}}\over m_c^2}\right)\Bigg\},   \non \\
\Gamma^{\rm ann}(\Xi_{cc}^+) &=& 3{G_F^2m_c^2\over \pi}\,|V_{cs}V_{ud}|^2\,
r_{_{\Xi_{cc}}}\left|\psi^{D}_{c\bar d}(0)\right|^2 \left((1-x)^2+\left|{V_{cd}\over V_{cs}}\right|^2\right)\Bigg\{\Big(\tilde B(c_1^2+c_2^2)-2c_1c_2\Big)  \non \\
&+& {1\over 2}\Big(\tilde \beta(c_1^2+c_2^2)-2c_1c_2\Big)\eta\left( {m^2_{\Xi_{cc}}-m^2_{\{cc\}}\over m_c^2} \right)\Bigg\},   \non \\
\Gamma^{\rm int}_+(\Omega_{cc}^+) &=& {G_F^2m_c^2\over 6\pi}\,|V_{cs}V_{ud}|^2\, r_{_{\Omega_{cc}}}\left|\psi^{D_s}_{c\bar s}(0)\right|^2
\Bigg\{ \Big(2c_1c_2+N_cc_1^2-\tilde B c_2^2\Big)\Big(5+\left|{V_{us}\over V_{ud}}\right|^2(1-x)^2(5+x)\Big)
\non \\
&-&  {9\over 2}\Big(2c_1c_2+N_cc_1^2-\tilde \beta c_2^2 \Big)\eta\left( {m^2_{\Omega_{cc}}-m^2_{\{cc\}}\over m_c^2} \right)\Bigg\},   \\
\Gamma^{\rm int}_+(\Xi_{cc}^+) &=& {G_F^2m_c^2\over 6\pi}\,|V_{cd}V_{ud}|^2\, r_{_{\Xi_{cc}}}\left|\psi^{D}_{c\bar q}(0)\right|^2
\Bigg\{ \Big(2c_1c_2+N_cc_1^2-\tilde B c_2^2\Big)\Big(5+\left|{V_{us}\over V_{ud}}\right|^2(1-x)^2(5+x)\Big)
\non \\
&-&  {9\over 2}\Big(2c_1c_2+N_cc_1^2-\tilde \beta c_2^2\Big)\Big(1+\left|{V_{us}\over V_{ud}}\right|^2(1-x)^2(1+x)\Big)\eta\left( {m^2_{\Xi_{cc}}-m^2_{\{cc\}}\over m_c^2} \right)\Bigg\},   \non \\
\Gamma^{\rm int}_-(\Xi_{cc}^{++}) &=& -{G_F^2m_c^2\over 6\pi}\,|V_{cs}V_{ud}|^2\, r_{_{\Xi_{cc}}}\left|\psi^{D}_{c\bar q}(0)\right|^2
\Bigg\{ \Big(\tilde Bc_1^2-2c_1c_2-N_cc_2^2\Big)\Big((1-x)^2(5+x)+\left|{V_{cd}\over V_{cs}}\right|^2
\non \\
&+&  \left|{V_{us}\over V_{ud}}\right|^2\sqrt{1-4x}\, \Big)
 -{9\over 2}\Big(\tilde \beta c_1^2-2c_1c_2-N_cc_2^2\Big)(1-x)(1+x-{2\over 3}x^2)\eta\left( {m^2_{\Xi_{cc}}-m^2_{\{cc\}}\over m_c^2} \right)\Bigg\},   \non
\en
and
\be \label{eq:SL67}
\Gamma^{\rm SL}_{int}(\Omega_{cc}^{+}) &=& {G_F^2m_c^2\over 6\pi}|V_{cs}|^2r_{_{\Omega_{cc}}}|\psi^{D_s}_{c\bar s}(0)|^2\left[ 5-{9\over 2}(1-{5\over 6}z^2+{1\over 3}z^3)\left({m_{\Omega_{cc}}^2-m_{\{cc\}}^2\over m_c^2}\right)\right], \non \\
\Gamma^{\rm SL}_{int}(\Xi_{cc}^+)
&=& {G_F^2m_c^2\over 6\pi}|V_{cd}|^2r_{_{\Xi_{cc}}}|\psi^{D}_{c\bar d}(0)|^2\left[ 5-{9\over 2}(1-{5\over 6}z^2+{1\over 3}z^3)\left({m_{\Xi_{cc}}^2-m_{\{cc\}}^2\over m_c^2}\right)\right].
\en
Except for the weak annihilation term, the expression of Pauli interference will be very lengthy if the hadronic parameters $\eta^q_i$ and $\tilde \beta^q_i$ are all treated to be different from each other. Since in realistic calculations we will set $\tilde \beta^q_i(\mu_h)=1$ under valence quark approximation and put $\eta^q_i$ to unity, we shall assume for simplicity that $\eta^q_i=\eta$ and $\tilde \beta^q_i=\tilde\beta$.

As far as the dimension-6 spectator effects are concerned,
we now compare our results Eqs. (\ref{eq:Spectorcharmbary}) and (\ref{eq:SL67}) with Eqs. (13) and (8) of \cite{Guberina}.
Since we are working at the $\mu=m_Q$ scale, we need to set the parameter $\kappa$ appearing in  \cite{Guberina}  to be unity. Noting that $r_{\B_{cc}}|\psi_D^{c\bar q}(0)|^2=|\psi_{\B_{cc}}^{dq}(0)|^2$ in our case, we see that $\Gamma^{\rm int}_+$ and $\Gamma^{\rm int}_-$ obtained by  Guberina, Meli\'c and H.~\v Stefan\v ci\'c (GMS) are larger than ours by a factor of 3/2, whereas their $\Gamma^{\rm ann}$ ($\Gamma^{\rm SL}$) is smaller than ours by a factor of 6/5 (2). Because the wave function of the doubly charmed baryon is related to that of the charmed meson through the relation $|\psi^{dq}(0)|^2={2\over 3}|\psi^{c\bar q}(0)|^2$ by GMS, it turns out that while we agree on the $\Gamma^{\rm int}_+$ and $\Gamma^{\rm int}_-$ in terms of $|\psi^{c\bar q}(0)|^2$, the expressions of  $\Gamma^{\rm ann}$ and $\Gamma^{\rm SL}_{int}$ by GMS are smaller than ours by a factor of 9/5 and 3, respectively.

\subsection{Numerical results}

To compute the decay widths of doubly charmed baryons,
we have to specify the values of $\tilde B$ and $r_{\B_{cc}}$. Since $\tilde B=1$ in the
valence-quark approximation and since the wavefunction squared ratio $r$
is evaluated using the quark model, it is reasonable to assume that the NQM
and the valence-quark approximation are most reliable when the baryon matrix
elements are evaluated at a typical hadronic scale $\mu_{\rm had}$. As
shown in \cite{Neubert97}, the parameters $\tilde B$ and $r$ renormalized
at two different scales are related via the renormalization group equation
to be
\be \label{eq:RGE1}
\tilde B(\mu)r(\mu) =\, \tilde B(\mu_{\rm had})r(\mu_{\rm had}),  \qquad \tilde B(\mu) =\, {\tilde{B}(\mu_{\rm had})\over \kappa+{1\over N_c}(\kappa
-1)\tilde B(\mu_{\rm had}) }\,,
\en
with
\be \label{eq:RGE2}
\kappa=\left({\alpha_s(\mu_{\rm had})\over \alpha_s(\mu)}\right)^{3N_c/2
\beta_0}=\sqrt{\alpha_s(\mu_{\rm had})\over \alpha_s(\mu)}
\en
and $\beta_0={11\over 3}N_c-{2\over 3}n_f$. The parameter $\kappa$ takes care of the evolution from $m_Q$ to the hadronic scale.
We consider the hadronic scale in the range of $\mu_{\rm had}\sim 0.65-1$ GeV. Taking  the  scale $\mu_{\rm had}=0.90$ GeV as an illustration, we obtain $\alpha_s(\mu_{\rm had})
=0.59$, $\tilde{B}(\mu)=0.75\tilde B(\mu_{\rm
had})\simeq 0.75$ and $r(\mu)\simeq 1.33\,r(\mu_{\rm had})$.
The parameter $\tilde\beta$ is treated in a similar way.

For numerical calculations, we use $c_1(\mu)=1.346$ and $c_2(\mu)=-0.636$ evaluated at the scale $\mu=1.25$ GeV with $\Lambda^{(4)}_{\overline {\rm MS}}=325$ MeV \cite{Buras}, $m_{\Xi_{cc}^*}-m_{\Xi_{cc}}=106$ MeV and $m_{\Omega_{cc}^*}-m_{\Omega_{cc}}=94$ MeV from \cite{Ebert:2005}, the wave function $|\psi^{cc}(0)|=0.17\,{\rm GeV}^{3/2}$ and the average kinetic energy $T=0.37$ GeV from \cite{Kiselev:1999}. For the decay constants, we use $f_D=204$ MeV and $f_{D_s}=250$ MeV.
For the charmed quark mass we use $m_c=1.56$ GeV fixed from the experimental values for $D^+$ and $D^0$ semileptonic widths \cite{Cheng:2018}.

The results of calculations to order $1/m_c^3$ are exhibited in Table \ref{tab:lifetime3_charmbary}.
The lifetime hierarchy  $\tau(\Xi_{cc}^{++})>\tau(\Omega_{cc}^+)>\tau(\Xi_{cc}^+)$ is  understandable. The $\Xi_{cc}^{++}$ baryon is
longest-lived owing to the destructive
Pauli interference, while $\Xi_{cc}^+$ is shortest-lived due to the fact that $\Gamma^{\rm ann}(\Xi_{cc}^+)\gg\Gamma^{\rm int}_+(\Omega_{cc}^+)$. From Eq. (\ref{eq:Spectorcharmbary}) we see that apart from QCD corrections to Wilson coefficients, $\Gamma^{\rm ann}/\Gamma^{\rm int}_+$ is basically of order 18/5. As for semileptonic decay rates, we have $\Gamma^{\rm SL}(\Omega_{cc}^+)\gg \Gamma^{\rm SL}(\Xi_{cc}^+)>\Gamma^{\rm SL}(\Xi_{cc}^{++})$ owing to a large Pauli interference effect in the $\Omega_{cc}^+$ but Cabibbo-suppressed in the $\Xi_{cc}^+$. We have checked the lifetimes of doubly charmed baryons against the hadronic scale $\mu_{\rm had}$. The $\Xi_{cc}^{++}$ lifetime remains nearly constant, $\tau(\Xi_{cc}^+)$ is increased by 10\%, while $\tau(\Omega_{cc}^+)$ increased by 35\%
when the hadronic scale varies from 0.65 to 1.0 GeV.

\begin{table}
\caption{Various contributions to the decay rates (in units of
$10^{-12}$ GeV) of doubly charmed baryons to order $1/m_c^3$ with the hadronic scale $\mu_{\rm had}=0.90$ GeV.
}
\label{tab:lifetime3_charmbary}
\begin{center}
\begin{tabular}{|l c c c r r r  l c |} \hline \hline
 & $\Gamma^{\rm dec}$ & $\Gamma^{\rm ann}$ & $\Gamma^{\rm int}_-$ &
$\Gamma^{\rm int}_+$ & ~ $\Gamma^{\rm semi}$ & $\Gamma^{\rm tot}$~~ &
~$\tau(10^{-13}s)$~ & $\tau_{\rm expt}(10^{-13}s)$~  \\
\hline
 ~$\Xi_{cc}^{++}$ & ~~2.198 &  & $-1.383$ &  & ~0.450 &
~~~1.265~ & ~ 5.20~  & ~$2.56^{+0.28}_{-0.26}$~ \\
 ~$\Xi_{cc}^+$ & ~~2.198 & ~8.628~ & ~$$~ & ~~0.123 & ~0.525 &
~~~11.475~ & ~ 0.57~  &  \\
 ~$\Omega_{cc}^+$ & ~~2.148 & 0.611 & & $3.217$ & ~$2.445$ &
~~~8.421~ & ~ 0.78   & \\
\hline \hline
\end{tabular}
\end{center}
\end{table}

\begin{table}
\caption{Various contributions to the decay rates (in units of
$10^{-12}$ GeV) of doubly charmed baryons to order $1/m_c^4$ with the hadronic scale $\mu_{\rm had}=0.90$ GeV.
}
\label{tab:lifetime4_charmbary}
\begin{center}
\begin{tabular}{|l c r c r r r  c c |} \hline \hline
 & $\Gamma^{\rm dec}$ & $\Gamma^{\rm ann}$~~ & $\Gamma^{\rm int}_-$ &
$\Gamma^{\rm int}_+$ & ~ $\Gamma^{\rm semi}$ & $\Gamma^{\rm tot}$~~ &
~$\tau(10^{-13}s)$~ & $\tau_{\rm expt}(10^{-13}s)$~  \\
\hline
 ~$\Xi_{cc}^{++}$ & ~~2.198 &  & $-0.437$ &  & ~0.451 &
~~~2.212~ & ~ 2.98~  & $2.56^{+0.28}_{-0.26}$~ \\
 ~$\Xi_{cc}^+$ & ~~2.198 & ~~~12.260~ & ~$$~ & ~~0.030 & ~0.469 &
~~~14.958~ & ~ 0.44~ &   \\
 ~$\Omega_{cc}^+$ & ~~2.148 & ~0.979~ & & $-0.246$ & ~$0.318$ &
~~~3.200~ &  ~ 2.06~  &  \\
\hline \hline
\end{tabular}
\end{center}
\end{table}

As shown in \cite{Cheng:2018},
the heavy quark expansion in $1/m_c$ does not work well for describing the lifetime pattern of singly charmed baryons. Since the charm quark is not heavy enough, it is sensible to consider the subleading $1/m_c$ corrections to spectator effects as depicted in Eq. (\ref{eq:Spectorcharmbary}). The numerical results are shown in Table \ref{tab:lifetime4_charmbary}. By comparing Table \ref{tab:lifetime4_charmbary} with Table \ref{tab:lifetime3_charmbary}, we see that the lifetimes of $\Xi_{cc}^{++}$ and $\Xi_{cc}^+$ become shorter, while $\tau(\Omega_{cc}^+)$ becomes longer. This is because $\Gamma^{int}_+$ and $\Gamma^{semi}$ for $\Omega_{cc}^+$ are subject to large cancellation between dimension-6 and -7 operators. Such cancellation also occurs in $\Xi_{cc}^+$ but not so dramatic as the constructive Pauli interference there is Cabibbo-suppressed. We see from Table \ref{tab:lifetime4_charmbary} that $\Gamma^{int}_+(\Omega_{cc}^+)$ even becomes negative. This is because the dimension-7 contribution $\Gamma^{\rm int}_{+,7}(\Omega_{cc}^+)$ is destructive and its size are so large that it overcomes the dimension-6 one and flips the sign. This implies that the subleading corrections are too large to justify the validity of the HQE.

\begin{table}[t]
\caption{Various contributions to the decay rates (in units of
$10^{-12}$ GeV) of the $\Omega_{cc}^+$ after including subleading $1/m_c$ corrections to spectator effects. However, the dimension-7 contributions $\Gamma^{\rm int}_{+,7}$ and $\Gamma^{\rm SL}_7$ are multiplied by a factor of $(1-\alpha)$ with $\alpha$ varying from 0 to 1. }
\label{tab:lifetime4_charmbary_1}
\begin{center}
\begin{tabular}{|c c c  r r l  l  |} \hline \hline
$\alpha$ & $\Gamma^{\rm dec}$ & $\Gamma^{\rm ann}$ &
$\Gamma^{\rm int}_+$ & ~ $\Gamma^{\rm semi}$ & ~~~~$\Gamma^{\rm tot}$ &
~$\tau(10^{-13}s)$~  \\
\hline
 ~0~~ & ~~2.148 & ~0.979~ & $-0.246$ & ~$0.318$ &
~~~3.200~ & ~ 2.06   \\
 ~0.08~~ & ~~2.148 & ~0.979~ & $0.031$ & ~$0.489$ &
~~~3.647~ & ~ 1.80   \\
 ~0.30~~ & ~~2.148 & ~0.979~ & $0.792$ & ~$0.956$ &
~~~4.876~ & ~ 1.35   \\
 ~1~~ & ~~2.148 & ~0.979~ & $3.217$ & ~$2.445$ &
~~~8.789~ & ~ 0.75   \\
\hline \hline
\end{tabular}
\end{center}
\end{table}

In order to allow a description of the $1/m_c^4$ corrections to $\Gamma(\Omega_{cc}^+)$ within the realm of perturbation theory, we follow \cite{Cheng:2018} to introduce a  parameter $\alpha$ so that $\Gamma^{\rm int}_{+,7}(\Omega_{cc}^+,\Xi_{cc}^+)$ and $\Gamma^{\rm SL}_7(\Omega_{cc}^+,\Xi_{cc}^+)$ are multiplied by a factor of $(1-\alpha)$; that is, $\alpha$ describes the degree of suppression. In Table \ref{tab:lifetime4_charmbary_1} we show the variation of the $\Omega_{cc}^+$ lifetime with $\alpha$. At $\alpha=0.08$, $\Gamma^{\rm int}_{+}(\Omega_{cc}^+)$ starts to become positive, where $\tau(\Omega_{cc}^+)=1.80\times 10^{-13}s$. Since we do not know what is the value of $\alpha$, we can only conjecture that
the $\Omega_{cc}^+$ lifetime lies in the range
\be
0.75\times 10^{-13}s<\tau(\Omega_{cc}^+)< 1.80\times 10^{-13}s.
\en
For the $\Xi_{cc}^+$, its lifetime is rather insensitive to the variation of $\alpha$ as both  $\Gamma^{\rm int}_{+,7}(\Xi_{cc}^+)$ and $\Gamma^{\rm SL}_7(\Xi_{cc}^+)$ are Cabibbo-suppressed.

Our prediction of $\tau(\Xi_{cc}^{++})$ is slightly larger than the LHCb measurement (\ref{eq:LHCbtauXiccpp}).
We learn from \cite{Cheng:2018} that the predicted lifetimes of heavy mesons or baryons are always longer than the measured values. Presumably, this is because we have not yet taken into account all possible QCD corrections fully. Nevertheless, the lifetime ratios should be more trustworthy than the absolute lifetimes themselves. In the present work, we find that the ratio $\tau(\Xi_{cc}^{++})/\tau(\Xi_{cc}^+)$ is $\sim$ 9.1 to order $1/m_c^3$ and $\sim$ 6.7 to order $1/m_c^4$.

\section{Conclusions}
In this work we have analyzed the lifetimes of doubly charmed hadrons within the framework of the heavy quark expansion. It is well known that the lifetime differences stem from spectator effects such as $W$-exchange and Pauli interference.  We rely on the quark model to evaluate the hadronic matrix elements of dimension-6 and -7 four-quark operators responsible for spectator effects.

\vskip 0.2cm
The main results of our analysis are as follows.
\begin{itemize}

\item The doubly charmed baryon matrix element of the $\sigma\cdot G$ operator receives three distinct contributions: the interaction of the heavy quark with the chromomagnetic field produced from the light quark and from the other heavy quark, and the so-called Darwin term in which the heavy quark interacts with the chromoelectric field. The last term arises because the appropriate theory for dealing hadrons with more than one heavy quark is NRQCD rather than HQET.

\item  The $\Xi_{cc}^{++}$ baryon is longest-lived in the doubly charmed baryon system owing to the destructive Pauli interference absent in the $\Xi_{cc}^+$ and $\Omega_{cc}^+$. In the presence of dimension-7 contributions, its lifetime is reduced from $\sim5.2\times 10^{-13}s$ to $\sim3.0\times 10^{-13}s$.

\item The $\Xi_{cc}^{+}$ baryon has the shortest lifetime of order $0.45\times 10^{-13}s$ due to a large contribution from the $W$-exchange box diagram.

\item It is difficult to make a precise statement on the lifetime of $\Omega_{cc}^+$. Contrary to $\Xi_{cc}$ baryons, $\tau(\Omega_{cc}^+)$ becomes longer in the presence of dimension-7 effects so that the Pauli interference $\Gamma^{\rm int}_+$ even becomes negative. This means that the subleading corrections are too large to justify the validity of the HQE. Demanding the rate $\Gamma^{\rm int}_+$ to be positive for a sensible HQE, we conjecture that the $\Omega_c^0$ lifetime lies in the range of $(0.75\sim 1.80)\times 10^{-13}s$.

 \item The lifetime hierarchy pattern is  $\tau(\Xi_{cc}^{++})>\tau(\Omega_{cc}^+)>\tau(\Xi_{cc}^+)$ and the lifetime ratio $\tau(\Xi_{cc}^{++})/\tau(\Xi_{cc}^+)$ is predicted to be of order 6.7.

\end{itemize}

\section{Acknowledgments}
We would like to thank Robert Shrock for helpful discussions. One of
us (H.Y.C.) wishes to thank the hospitality of the C.N. Yang Institute for Theoretical Physics, Stony Brook University.
This research of H.Y.C. and Y.L.S.  was supported in part by the Ministry of Science and Technology of R.O.C. under Grant No. 106-2112-M-001-015 and U.S. National
Science Foundation Grant NSF-PHY-16-1620628, respectively.

%
%
\newcommand{\bi}{\bibitem}
%

\end{document}